\documentclass[conference]{IEEEtran}
\usepackage{amsmath,amsfonts}
\usepackage{graphicx,amssymb}
\usepackage{algorithmic,algorithm}
\makeatletter
\newcommand{\removelatexerror}{\let\@latex@error\@gobble}
\makeatother
\usepackage{amsthm,dsfont}
\usepackage{subfigure,url}
\usepackage[noadjust]{cite}
\usepackage{xcolor}%定义了一些颜色
\usepackage{colortbl,booktabs}%第二个包定义了几个*rule
\usepackage{multirow}
\theoremstyle{plain}

\theoremstyle{remark}

\newcounter{longequ}[longequ]
\addtocounter{longequ}{11}

\allowdisplaybreaks

\begin{document}
% \title{FDD Massive MIMO CSI Feedback with Lightweight and Flexible Deep Equilibrium Learning}
\title{Lightweight and Flexible Deep Equilibrium Learning for CSI Feedback in FDD Massive MIMO
}
\author{\IEEEauthorblockN{Yifan Ma\IEEEauthorrefmark{1}, Wentao Yu\IEEEauthorrefmark{1},
		Xianghao Yu\IEEEauthorrefmark{2}, Jun Zhang\IEEEauthorrefmark{1}, Shenghui Song\IEEEauthorrefmark{1}, and Khaled B. Letaief\IEEEauthorrefmark{1}}\\
	\IEEEauthorblockA{\IEEEauthorrefmark{1}Dept. of ECE, The Hong Kong University of Science and Technology, Hong Kong\\
	\IEEEauthorrefmark{2}Dept. of EE, City University of Hong Kong, Hong Kong\\}
	Email: \IEEEauthorrefmark{1}\{ymabj, wyuaq, eejzhang, eeshsong, eekhaled\}@ust.hk, \IEEEauthorrefmark{2}alex.yu@cityu.edu.hk
}
\maketitle

\begin{abstract}
In frequency-division duplexing (FDD) massive multiple-input multiple-output (MIMO) systems, downlink channel state information (CSI) needs to be sent back to the base station (BS) by the users, which causes prohibitive feedback overhead. In this paper, we propose a lightweight and flexible deep learning-based CSI feedback approach by capitalizing on deep equilibrium models. Different from existing deep learning-based methods that stack multiple explicit layers, we propose an implicit equilibrium block to mimic the behavior of an infinite-depth neural network. In particular, the implicit equilibrium block is defined by a fixed-point iteration and %the forward fixed-point finding procedure is separated from the backward neural network training. 
the trainable parameters in different iterations are shared, which results in a lightweight model. Furthermore, the number of forward iterations can be adjusted according to users' computation capability, enabling a flexible accuracy-efficiency trade-off. Simulation results will show that the proposed design obtains a comparable performance as the benchmarks but with much-reduced complexity and permits an accuracy-efficiency trade-off at runtime.% no or little performance loss. the superiority of the proposed method over various benchmarks in terms of both the spectral efficiency and scalability in large-scale wireless networks.
\end{abstract}
%%\begin{IEEEkeywords}
%%Intelligent reflecting surfaces, large-scale optimization, gradient descent
%%\end{IEEEkeywords}

\IEEEpeerreviewmaketitle
\section{Introduction}
Massive multiple-input multiple-output (MIMO) systems are regarded as a key enabler for 5G and beyond wireless communication systems \cite{FiveG} where frequency-division duplexing (FDD) is considered a compelling operation mode. In FDD massive MIMO systems, users need to feed the downlink channel state information (CSI) back to the base station (BS) to facilitate beamforming. However, the dimension of the CSI increases significantly as the number of antennas at the BS gets larger, which causes prohibitive feedback overhead. Conventional compressive sensing (CS)-based methods were widely applied for CSI compression and recovery \cite{Sparsity}, but they suffered from noteworthy limitations, such as the impractical assumption of channel sparsity, the limited ability to exploit the channel structures, and the high computational cost of the iterative operations \cite{CsiNet}.
% Note that the memory and computational resources of mobile devices are limited, which poses challenges to the deployment of complicated feedback algorithms. 
% Moreover, the computational budget varies (e.g., numerous background applications reduce the available computation capability) and the energy budget varies (e.g., a mobile phone may be in power-saving mode) in practice.
% Therefore, it is of vital importance to design a lightweight and adaptive CSI feedback scheme for FDD massive MIMO systems, permitting instant accuracy-efficiency trade-off at runtime.
% Therefore, it is of vital importance to design an effective and efficient CSI feedback scheme for FDD massive MIMO systems.

%To effectively compress the downlink CSI, traditional compressive sensing (CS) methods are widely applied \cite{Sparsity, CS}. However, these CS-based methods have noteworthy limitations, such as the ideal assumption of channel sparsity, lack full utilization of channel structures, and the high computational cost of the iterative process \cite{CsiNet}. 
Thanks to the the universal approximation capability of neural networks, the deep learning-based auto-encoder and decoder structures have been leveraged to effectively compress and reconstruct the downlink CSI. The exploratory work \cite{CsiNet} proposed a convolutional neural network (CNN)-based CsiNet which outperforms the CS-based algorithms especially with low compression ratios. Several subsequent studies, including ConvCsiNet \cite{ConvCsiNet} and TransNet \cite{TransNet}, aimed to further improve the feedback accuracy using deeper CNNs and attention mechanism, respectively. However, the performance improvement is at the cost of computational complexity. For example, the number of floating point operations (FLOPs) of ConvCsiNet is almost one hundred times that of CsiNet. % Since the memory and computational resources of mobile devices are limited, the high memory and computational cost makes the deployment of these deep learning-based designs challenging in practice.
This puts prohibitive burdens on memory and computational resources, making the deployment of these deep learning-based designs challenging in practice.

% To overcome the huge execution cost, various lightweight feedback schemes are proposed. On the one hand, neural network compression techniques are adopted to reduce the neural network complexity. Specifically, the neural network weight pruning and binarization were introduced in \cite{NNCompression}, which greatly save the memory with slight performance loss. Knowledge distillation was utilized in \cite{Knowledge} where the knowledge of a complex model is transferred to a simple model so that the performance of the lightweight model is improved. On the other hand, efficient neural network architecture designs also attract much attention. For example, the vanilla convolutional layers were replaced with shuffle layers in \cite{ConvCsiNet}, which achieved a comparable accuracy when the number of FLOPs is 1/3 of ConvCsiNet. However, the computational budget varies (e.g., numerous background applications reduce the available computation capability) and the energy budget differs (e.g., a mobile phone may be in power-saving mode) in practice. Different models need to be allocated to different situations according to resource budget and the delay for downloading data and freeing-up spaces is not negligible. The trade-off between accuracy and effciency is thus inflexible and has many constraints for existing methods. 
To reduce the huge execution cost, various lightweight schemes were proposed. Efficient neural network architectures were designed by replacing redundant neural network layers with simplified compositions. For example, the vanilla convolutional layers were replaced by shuffle layers in \cite{ConvCsiNet} to achieve a comparable accuracy with 1/3 FLOPs of ConvCsiNet. However, existing models can not fit the dynamic communication environment. In practice, the computational budget varies (e.g., numerous background applications can reduce the available computational resource) and the energy budget differs (e.g., a mobile phone may be in the power-saving mode). As such, different models need to be pre-trained, stored, and allocated to different devices/conditions according to the dynamic resource budget, leading to large memory and time cost. Hence, it is of great value to develop new models that provide a flexible trade-off between accuracy and efficiency. % Besides, the delay for downloading data and freeing-up spaces is not negligible. 

In this paper, we propose a lightweight and flexible CSI feedback approach for FDD massive MIMO systems, allowing an accuracy-efficiency trade-off at runtime. Instead of stacking multiple well-designed \emph{explicit} layers, i.e., the output is computed by explicitly cascading a series of nonlinear mappings, we develop a learning model composed of \emph{implicit} equilibrium blocks. Specifically, the equilibrium block specifies the input-output relationship by a fixed-point equation. The forward propagation is a fixed-point finding process, which is separated from the backward neural network training, keeping a constant memory consumption of the backpropagation \cite{bai2019DEQ}. Besides, the trainable parameters in different iterations are shared, which results in a lightweight model. To meet the dynamically changing computation capability in practice, %we adopt a single-stream implicit learning model at the user side and a multiscale implicit learning model at the BS \cite{bai2020MDEQ}. 
the number of implicit iterations at the users and the BS can be adjusted according to latency and computational budget during the inference stage. % At the BS side, we apply an efficient solver to generate the output of the equilibrium block, which is equivalent to running an infinite-depth explicit neural network. 
Extensive simulation results show that the proposed method outperforms the conventional CsiNet \cite{CsiNet} and has a comparable or even better performance than the existing approaches \cite{ConvCsiNet} with greatly reduced complexity.

\section{System Model and Problem Formulation}\label{sec:sys}

This work considers a single-cell FDD massive MIMO system where the BS is equipped with $N_t$ transmit antennas and the user is equipped with a single receive antenna. For ease of illustration, a single user case is considered while the proposed scheme can be easily generalized to the multi-user scenario. An orthogonal frequency division multiplexing (OFDM) system with $N_c$ subcarriers is considered. The received signal on the $n$-th subcarrier is expressed as
\begin{equation}
y_n = \mathbf{h}_n^H \mathbf{v}_n x_n + z_n,
\end{equation}
where $\mathbf{h}_n \in \mathbb{C}^{N_t \times 1}$, $\mathbf{v}_n \in \mathbb{C}^{N_t \times 1}$, $x_n \in \mathbb{C}$, and $z_n \in \mathbb{C}$ denote the downlink channel vector, the downlink beamforming vector, the transmit symbol, and the additive noise of the $n$-th subcarrier, respectively. The CSI matrix over all subcarriers is thus denoted by $\mathbf{H} = [\mathbf{h}_1, \cdots, \mathbf{h}_{N_c}]^H \in \mathbb{C}^{N_c \times N_t}$. The downlink beamforming design requires the BS to know the downlink CSI. In this paper, we assume that the downlink channel is perfectly known at the user side via pilot-based training and focus on the efficient feedback design \cite{CsiNet, ConvCsiNet, TransNet}.
%Due to the large number of complex elements in $\mathbf{H}$, CSI feedback may cause severe delay and overhead in practice.  

Considering that the channel matrix $\mathbf{H}$ contains $2 N_c N_t$ real elements and the feedback overhead is prohibitive for FDD massive MIMO system, we first sparsify $\mathbf{H}$ in the angular-delay domain using a 2D discrete Fourier transform (2D-DFT) \cite{CsiNet} as follows
\begin{equation}
\mathbf{H}' = \mathbf{F}_\mathrm{d} \mathbf{H} \mathbf{F}_\mathrm{a},
\end{equation}
where $\mathbf{F}_\mathrm{d} \in \mathbb{C}^{N_c \times N_c}$ and $\mathbf{F}_\mathrm{a} \in \mathbb{C}^{N_t \times N_t}$ are two DFT matrices. Only the first $N_a$ rows of $\mathbf{H}'$ contain significant values and other elements are close to zero because the time delays between multipath arrivals are within a limited period \cite{CsiNet}. Therefore, we take the first $N_a$ rows of $\mathbf{H}'$ ($N_a < N_c$) and define a new matrix $\mathbf{H}'' \in \mathbb{C}^{N_a \times N_t}$. By doing this, we can compress $\mathbf{H}''$ instead of $\mathbf{H}$ with only $2N_a N_t$ entries and imperceptible information loss. 

In this work, a deep learning-based method is proposed for CSI compression and recovery. The encoding process at the user side is given by
\begin{equation} \label{encoder}
\mathbf{s} = \mathcal{E}_{\theta_\mathrm{e}}(\mathbf{H}''),
\end{equation}
which further compresses the channel matrix into an $M \times 1$ codeword $\mathbf{s}$. The parameterized mapping $\mathcal{E}_{\theta_\mathrm{e}}(\cdot)$ denotes the compression procedure and $\theta_\mathrm{e}$ is the trainable parameters in the encoder. The compression ratio is defined as $\gamma = M/2N_a N_t$. 
We use the same setting as \cite{CsiNet, ConvCsiNet, TransNet} and assume $\mathbf{s}$ is sent back to the BS via error-free transmission.
After receiving the codeword, the BS reconstructs the channel matrix through a decoder, expressed as
\begin{equation} \label{decoder}
\hat{\mathbf{H}}'' = \mathcal{D}_{\theta_\mathrm{d}}(\mathbf{s}),
\end{equation}
where $\mathcal{D}_{\theta_\mathrm{d}}(\cdot)$ denotes the recovery procedure and $\theta_\mathrm{d}$ represents the trainable parameters at the decoder. The objective is to minimize the mean-squared-error (MSE) between the recovered channel and the true channel, given by
%\begin{equation}
%\min_{\theta_\mathrm{e}, \theta_\mathrm{d}} \quad \mathbb{E} ||\mathbf{H}'' - \hat{\mathbf{H}}''||_2^2 = \mathbb{E} ||\mathbf{H}'' - \mathcal{D}_{\theta_\mathrm{d}}(\mathcal{E}_{\theta_\mathrm{e}}(\mathbf{H}''))||_2^2.
%\end{equation}
\begin{equation}
\min_{\theta_\mathrm{e}, \theta_\mathrm{d}} \quad \mathbb{E} \left\{||\mathbf{H}'' - \mathcal{D}_{\theta_\mathrm{d}}(\mathcal{E}_{\theta_\mathrm{e}}(\mathbf{H}''))||_2^2 \right\}.
\end{equation}
The mapping $\mathcal{E}_{\theta_\mathrm{e}}(\cdot)$ and $\mathcal{D}_{\theta_\mathrm{d}}(\cdot)$ can be instantiated as an auto-encoder and a decoder, and jointly trained via end-to-end learning \cite{CsiNet, ConvCsiNet, TransNet}. However, most of the existing works improve the reconstruction accuracy at the cost of high neural network complexity, which is not affordable for mobile devices due to limited resources. Moreover, the existing feedback schemes lack an effective mechanism to achieve the performance-complexity trade-off during execution. In practice, different devices have different runtime for a neural network \cite{AIBenchmark}. Given the same latency budget, high-end devices can achieve better performance by running complicated models, while low-end ones have to sacrifice accuracy to meet the response time constraint. Even for the same device, the resource availability varies under different situations. Therefore, it is of vital importance to design a lightweight and flexible method for CSI feedback. % Therefore, a lightweight and adaptive CSI feedback scheme is of vital importance.

\section{Proposed CSI Feedback Scheme with Deep Equilibrium Learning}
In this section, we first present the overall diagram of the proposed deep equilibrium learning-based CSI feedback scheme and the delicate design of the encoding and decoding blocks. As two key advantages, we then present the lightweight and flexible inference procedure and a low-complexity training procedure. 
\subsection{Overall Structure}
The widely adopted explicit deep neural network model can be written as
\begin{equation}
\mathbf{y}_{\mathrm{out}}^{[i+1]} = f_\theta^{[i+1]} (\mathbf{y}_{\mathrm{out}}^{[i]}; \mathbf{x}_{\mathrm{in}}), \quad i = 0, 1, \cdots, L-1,
\end{equation}
where $\mathbf{y}_{\mathrm{out}}^{[i]}$ is the output of the $i$-th layer, $f_\theta^{[i]}$ represents the $i$-th layer's parameterized function, $\mathbf{x}_{\mathrm{in}}$ denotes the input, and $L$ is the number of layers. Recent findings in \cite{weighttied} showed that employing the same transformation function in each layer still leads to competitive results. The shared-weight neural network is correspondingly expressed as
\begin{equation}
\mathbf{y}_{\mathrm{out}}^{[i+1]} = f_\theta (\mathbf{y}_{\mathrm{out}}^{[i]}; \mathbf{x}_{\mathrm{in}}), \quad i = 0, 1, \cdots, L-1,
\end{equation}
and thus the number of trainable parameters is greatly reduced. Note that stacking an infinite number of weight-sharing layers corresponds to finding the fixed-point solution of the equation \cite{bai2019DEQ, Wentao22DEQ}
% Inspired by the fact that stacking infinite number of shared-weight layers corresponds to finding the fixed-point solution of the equation \cite{bai2019DEQ, Wentao22DEQ}
\begin{equation}
\mathbf{y}_{\mathrm{out}}^{*} = f_\theta (\mathbf{y}_{\mathrm{out}}^{*}; \mathbf{x}_{\mathrm{in}}).
\end{equation}
The implicit equilibrium model allows us to directly find the equilibrium point $\mathbf{y}_{\mathrm{out}}^{*}$ with an off-the-shelf solver or iteratively executing $f_\theta(\cdot)$. Besides, according to the implicit function theorem, the backward propagation is independent of the forward fixed-point finding process, making the trainable weight update equivalent to one single layer. % These advantages of deep equilibrium learning inspire multiple applications including natural language processing \cite{bai2019DEQ} and computer vision \cite{bai2020MDEQ}.

\begin{figure}[htbp] %H为当前位置，!htb为忽略美学标准，htbp为浮动图形
\centering
\includegraphics[width=0.49\textwidth]{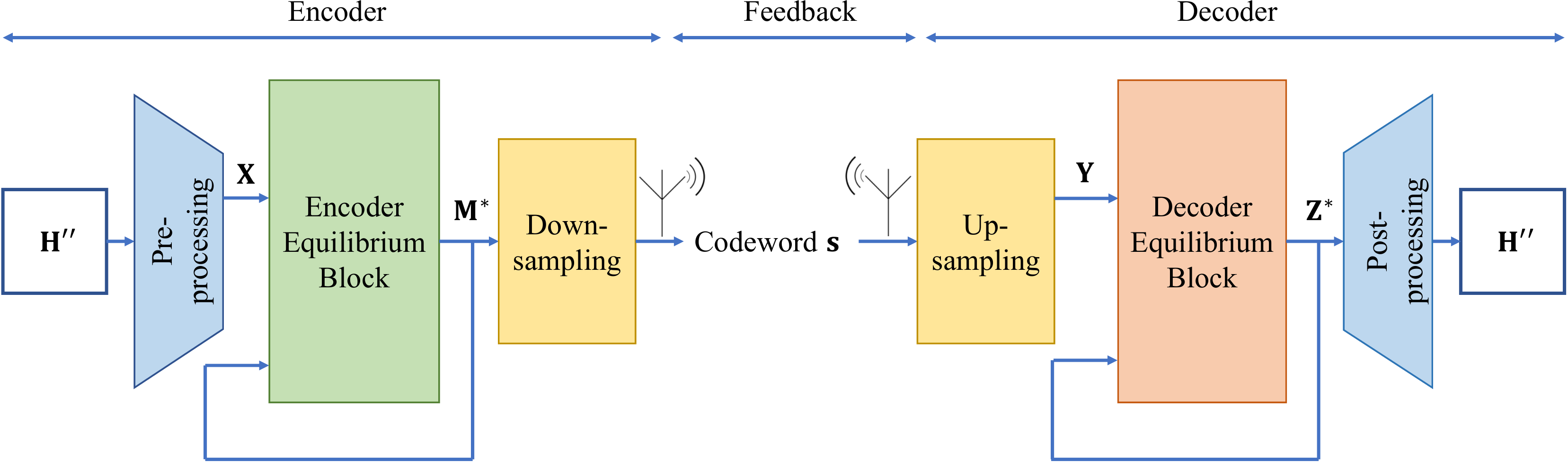} %插入图片，[]中设置图片大小，{}中是图片文件名
\caption{Diagram of the proposed deep equilibrium learning-based CSI feedback scheme.} %最终文档中希望显示的图片标题
\label{architecture} %用于文内引用的标签
\end{figure}

With the same trainable parameters in each iteration, the output and all hidden units of conventional deep equilibrium learning should have the same dimensions.
Nevertheless, for the CSI compression and recovery task, the input channel matrix needs to be downsampled and upsampled to guarantee $M \ll 2N_a N_t$. Hence, it is not feasible if we only apply equilibrium blocks for the mappings in \eqref{encoder} and \eqref{decoder}. Furthermore, the computation capability of mobile users is generally limited, which requires a simple yet effective encoder to be deployed at the user side. Based on these two aspects, we incorporate the implicit equilibrium block with traditional explicit neural network design. The overall diagram is shown in Fig. \ref{architecture} and the operations at the test stage are summarized in \textbf{Algorithm \ref{alg1}}. 

\begin{algorithm}[htbp]
\caption{Inference Stage of the Proposed CSI Feedback Scheme with Deep Equilibrium Learning}
\label{alg1}
\begin{algorithmic}[1]
\STATE {{\bf{Input:}} Truncated channel matrix in angular-delay domain $\mathbf{H}''$. % Maximum iteration number $T_{\mathrm{e}}$, $T_{\mathrm{d}}$ of the encoding and decoding blocks, respectively. 
Well-trained encoding blocks $f_{\mathrm{pre}}$, $f_{\mathrm{eim}}$, and $f_{\mathrm{down}}$. The FLOPs number of encoding blocks $F_{\mathrm{pre}}$, $F_{\mathrm{eim}}$, and $F_{\mathrm{down}}$. Well-trained decoding blocks $f_{\mathrm{up}}$, $f_{\mathrm{dim}}$, and $f_{\mathrm{post}}$. The FLOPs number of decoding blocks $F_{\mathrm{up}}$, $F_{\mathrm{dim}}$, and $F_{\mathrm{post}}$. Computational budget $R_\mathrm{e}$ and $R_\mathrm{d}$ at the user side and the BS side, respectively.} 
\STATE {{\bf{Output:}} Reconstructed channel matrix $\hat{\mathbf{H}}''$} 
%\STATE {{\bf{Initialize:}} $\mathbf{M}^{[0]} \leftarrow \mathbf{0}$ and $\mathbf{Z}^{[0]} \leftarrow \mathbf{0}$} 
\STATE {{\bf{Encoder:}}}
\STATE {$\mathbf{X} \leftarrow f_{\mathrm{pre}}(\mathbf{\mathbf{H}''})$}
\STATE {$\mathbf{M}^{[0]} \leftarrow \mathbf{X}$}
\STATE {$T_{\mathrm{e}} \leftarrow \lfloor \frac{R_\mathrm{e} - F_{\mathrm{pre}} - F_{\mathrm{down}}}{F_{\mathrm{eim}}} \rfloor$}
\FOR {$t \leftarrow 1$ to $T_{\mathrm{e}}$} 
\STATE {$\mathbf{M}^{[t]} \leftarrow f_{\mathrm{eim}}(\mathbf{M}^{[t-1]}, \mathbf{X})$}
% \STATE {$t \leftarrow t + 1$;}
\ENDFOR
% \UNTIL{$t> T_{\mathrm{e}}$} 
\STATE {$\mathbf{s} \leftarrow f_{\mathrm{down}}(\mathbf{M}^{[T_{\mathrm{e}}]})$}
\STATE {{\bf{Decoder:}}}
\STATE {$\mathbf{Y} \leftarrow f_{\mathrm{up}}(\mathbf{s})$}
\STATE {$\mathbf{Z}^{[0]} \leftarrow \mathbf{Y}$}
\STATE {$T_{\mathrm{d}} \leftarrow \lfloor \frac{R_\mathrm{d} - F_{\mathrm{up}} - F_{\mathrm{post}}}{F_{\mathrm{dim}}} \rfloor$}
% \REPEAT 
\FOR {$t \leftarrow 1$ to $T_{\mathrm{d}}$} 
\STATE {$\mathbf{Z}^{[t]} \leftarrow f_{\mathrm{dim}}(\mathbf{Z}^{[t-1]}, \mathbf{Y})$}
% \STATE {$t \leftarrow t + 1$;}
\ENDFOR
\STATE {$\hat{\mathbf{H}}'' \leftarrow f_{\mathrm{post}}(\mathbf{Z}^{[T_{\mathrm{d}}]}$)}
% \UNTIL{$t>T_{\mathrm{d}}$} 
%\STATE {$\hat{\mathbf{H}}'' \triangleq \mathbf{Z}^{*} \leftarrow g_\mathrm{solv}(f_{\mathrm{dim}}, \mathbf{Z}^{[0]}, \epsilon_{\mathrm{d}}, T_{\mathrm{d}})$;}
\end{algorithmic}
\end{algorithm}

Mathematically, the encoding process is expressed as
\begin{equation}
\begin{aligned}
& \mathbf{X} = f_{\mathrm{pre}}(\mathbf{\mathbf{H}''}), \\
& \mathbf{M}^{[t]} = f_{\mathrm{eim}}(\mathbf{M}^{[t-1]}, \mathbf{X}), \quad t = 1, \cdots, T_{\mathrm{e}},\\
& \mathbf{s} = f_{\mathrm{down}}(\mathbf{M}^{*}).
\end{aligned}
\end{equation}
$f_{\mathrm{pre}}(\cdot)$ is the preprocessing block to form the input injection $\mathbf{X}$. The input injection is pivotal to equilibrium models. Since the fixed-point $\mathbf{M}^{*}$ does not depend on any initial value of $\mathbf{M}^{[0]}$, preprocessing $\mathbf{\mathbf{H}''}$ and injecting $\mathbf{X}$ properly ensures the dependency between the fixed-point and the model input. $f_{\mathrm{eim}}(\cdot,\cdot)$ and $T_{\mathrm{e}}$ denote the encoder-side implicit equilibrium block and the number of iterations, respectively. 
% During the training stage, the forward fixed-point solution is calculated via an existing solver. 
% In the test stage, due to insufficient computational resources of devices, the implicit equilibrium block is terminated with an error threshold $\epsilon_{\mathrm{e}}$ and a maximum number of iterations $T_{\mathrm{e}}$, whichever reaches first. 
Note that the hyperparameter $T_{\mathrm{e}}$ can be adjusted flexibly during runtime according to the resource budget. Therefore, it provides us with a flexible CSI feedback approach applicable in different scenarios without additional training or model downloading costs. Details of flexible implementation will be discussed in Section \ref{flexibility}.
% $f_{\mathrm{down}}(\cdot)$ represents the downsampling module.
% The input first goes through a preprocessing block to form the input injection $\mathbf{x}$ into the encoder implicit block $f_{\mathrm{eim}}(\cdot,\cdot)$. 
The output of the implicit model $\mathbf{M}^{*}$ is then fed into the downsampling block $f_{\mathrm{down}}(\cdot)$ which reduces the output into $M$ dimensions. The codeword $\mathbf{s}$ is then transmitted back to the BS.

At the BS side, the decoding process is given by
\begin{equation}
\begin{aligned}
\mathbf{Y} &= f_{\mathrm{up}}(\mathbf{s}), \\
% & \mathbf{z}^{[t]} = f_{\mathrm{dim}}(\mathbf{z}^{[t-1]}, \mathbf{y}), \quad t = 1, \cdots, T_{\mathrm{d}}\\
\mathbf{Z}^{[t]} & = f_{\mathrm{dim}}(\mathbf{Z}^{[t-1]}, \mathbf{Y}), \quad t = 1, \cdots, T_{\mathrm{d}},\\
\hat{\mathbf{H}}'' & = f_{\mathrm{post}}(\mathbf{Z}^{*}),
\end{aligned}
\end{equation}
%\begin{IEEEeqnarray}{rCl}
%\mathbf{y} &=& f_{\mathrm{up}}(\mathbf{s}), \\
%% & \mathbf{z}^{[t]} = f_{\mathrm{dim}}(\mathbf{z}^{[t-1]}, \mathbf{y}), \quad t = 1, \cdots, T_{\mathrm{d}}\\
%\mathbf{z}^{*} & =& \underbrace{f_{\mathrm{dim}} \circ \cdots \circ f_{\mathrm{dim}}}_{\text{until convergence}}(\mathbf{z}^{[0]}, \mathbf{y}) \\
%& = &g_\mathrm{solv}(f_{\mathrm{dim}}, \mathbf{z}^{[0]}, \epsilon_{\mathrm{d}}, T_{\mathrm{d}}), \\
%\hat{\mathbf{H}}'' & =& \mathbf{z}^{*},
%\end{IEEEeqnarray}
where $f_{\mathrm{up}}(\cdot)$ represents the upsampling block, $\mathbf{Y}$ is the input injection of the decoder-side implicit equilibrium block $f_{\mathrm{dim}}(\cdot,\cdot)$, $\mathbf{Z}^{[0]}$ is the initial value, $T_{\mathrm{d}}$ denotes the maximum number of iterations, and $f_{\mathrm{post}}(\cdot)$ is the post-processing block. %The powerful computation capability of BS allows us to obtain a fixed-point transformation $f_{\mathrm{dim}}(\cdot,\cdot)$ with higher computational complexity than $f_{\mathrm{eim}}(\cdot,\cdot)$. 
The detailed structure of two equilibrium blocks will be discussed in the following subsection. % At both the training and inference stages, the output of the decoder-side implicit equilibrium module is computed by $g_\mathrm{solv}$, which will be discussed in the following subsection.
% After receiving the codeword, the codeword is firstly upsampled by $f_{\mathrm{up}}(\cdot)$ to form the input injection to the decoder implicit module $f_{\mathrm{eim}}(\cdot,\cdot)$, and the equilibrium point of this implicit module is then computed by an existing fixed point solver $g_\mathrm{solv}$ with a predetermined error tolerance $\epsilon_{\mathrm{d}}$ and a maximum number of iterations $T_{\mathrm{d}}$. The powerful processing units at the BS allow us to obtain an equilibrium with higher accuracy. 

%The encoding procedure include one preprocessing block denoted as $f_{\mathrm{pre}}(\cdot)$ is written as
%\begin{equation}
%\mathbf{s} = \mathcal{E}_{\theta_\mathrm{e}}(\mathbf{H}'') = f_{\mathrm{pre}}(\mathbf{H}''),
%\end{equation}

\subsection{Design of Encoding and Decoding Blocks}

\begin{figure}[t] %H为当前位置，!htb为忽略美学标准，htbp为浮动图形
\centering
\includegraphics[width=0.49\textwidth]{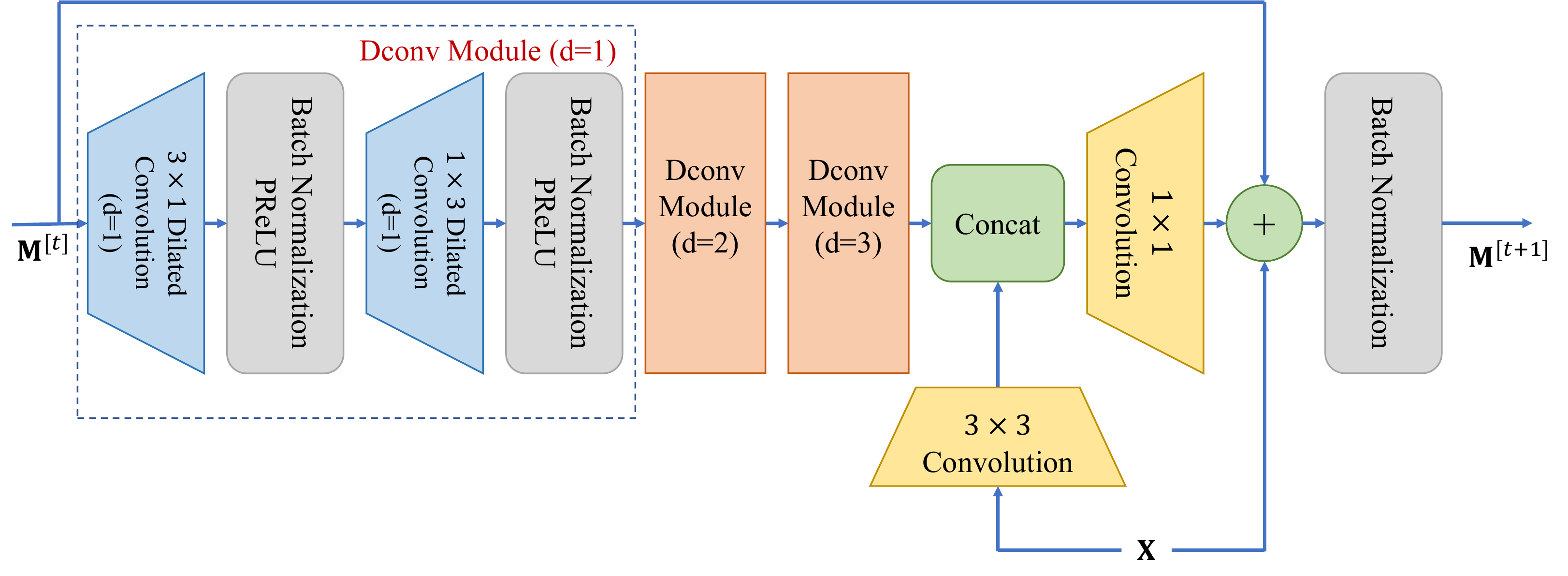} %插入图片，[]中设置图片大小，{}中是图片文件名
\caption{The implicit equilibrium block at the encoder.} %最终文档中希望显示的图片标题
\label{EnDe} %用于文内引用的标签
\end{figure}

\subsubsection{Encoder}
The input to the encoder is the real and the imaginary parts of $\mathbf{H}''$, forming a $2 \times N_a \times N_t$ dimensional tensor.
The preprocessing module of the encoder, i.e., $f_{\mathrm{pre}}(\cdot)$, is constructed by $5 \times 5$ convolutional kernels followed by batch normalization and parametric rectified linear unit (PReLU) activation functions. The PReLU function with a learnable parameter $\alpha$ is given by
\begin{equation}
\operatorname{PReLU}(x)= \begin{cases}x, & x \geq 0 \\ \alpha x, & x<0.\end{cases}
\end{equation}
This module effectively extracts the information from the input CSI matrix and fuses the features from both the real and the imaginary parts at an affordable cost. %The output of the preprocessing module is a $4 \times N_a \times N_t$ dimensional tensor. 
The output of the preprocessing module is treated as the input injection to the implicit equilibrium block. The major component of the equilibrium module is the transformation function $f_{\mathrm{eim}}(\cdot,\cdot)$. Inspired by the fact that the CNN-based autoencoder can efficiently extract the spatially local correlation in the CSI matrices \cite{CsiNet}, we develop the $f_{\mathrm{eim}}(\cdot,\cdot)$ by using a CNN-based structure, as shown in Fig. \ref{EnDe}. To improve the CSI reconstruction accuracy without increasing computational cost, dilated convolutions \cite{DCRNet} are adopted. % to increase the receptive field extract features with a specific interval $d$, which is also called as dilated rate. 
The 2D-dilated convolutional operation without bias is written as
\begin{equation}
(\mathbf{I} \circledast \mathbf{K})[i, j]=\sum_m \sum_n \mathbf{I}[i+d \times m, j+d \times n] \cdot \mathbf{K}[m, n],
\end{equation}
where $\circledast$, $d$, $\mathbf{I}$, and $\mathbf{K}$ denote the dilated convolution operator, dilated rate, input tensor, and convolutional kernel, respectively. $m$ and $n$ are the indices of convolutional kernel $\mathbf{K}$. % The effective kernel size of dilated convolution with dilated rate $d$ can be expressed as
%\begin{equation}
%k_i^{\prime}=k_i+\left(k_i-1\right)(d-1),
%\end{equation}
%where $k_i$ and $k_i^{\prime}$ are the used and effective kernel sizes of convolution, receptively.
When $d = 1$, the dilated convolution degenerates into the standard convolution. When $d > 1$, the dilated convolution operation provides a larger receptive field compared to the standard convolution with the same kernel size. %This helps achieve a sparse sampling for the block-sparse CSI matrix \cite{DCRNet}.
To promote good performance and diminish information loss, three consecutive dilated convolution modules are employed. In each dilated convolution module, $3 \times 1$ and $1 \times 3$ kernels are  applied to better extract the vertical and horizontal information.
Besides, inspired by the residual network \cite{he2016residual}, the identity shortcut connections of latent space variable $\mathbf{M}^{[t]}$ and input $\mathbf{X}$ are introduced. 
After the equilibrium module, the last downsampling block in the encoder is implemented by one fully-connected layer.

\subsubsection{Decoder}
\begin{figure}[t] %H为当前位置，!htb为忽略美学标准，htbp为浮动图形
\centering
\includegraphics[width=0.49\textwidth]{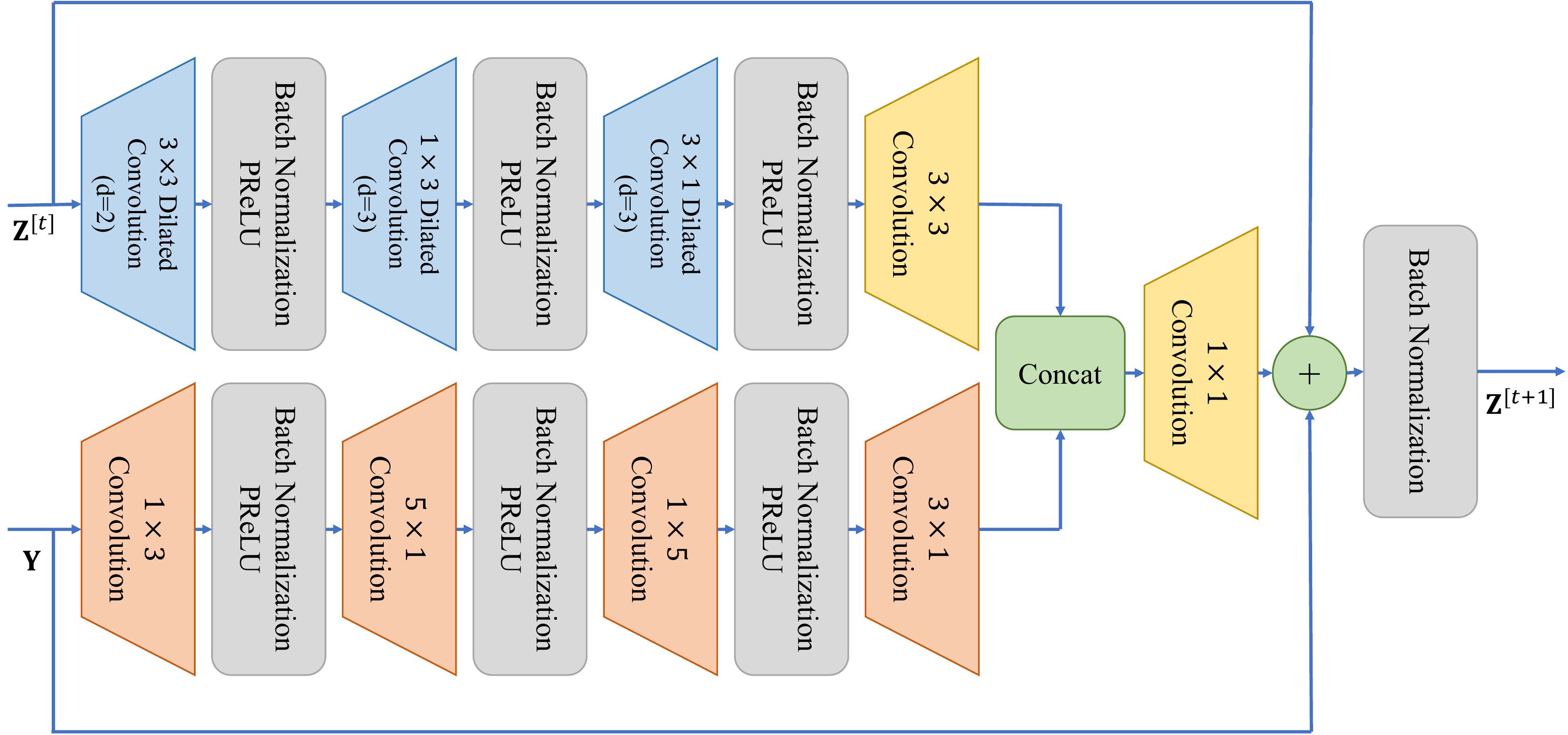} %插入图片，[]中设置图片大小，{}中是图片文件名
\caption{The implicit equilibrium block at the decoder.} %最终文档中希望显示的图片标题
\label{EnDe} %用于文内引用的标签
\end{figure}
After receiving the codeword $\mathbf{s}$, a fully-connected layer is adopted as $f_{\mathrm{up}}(\cdot)$ to recover the dimension of CSI matrix, forming the input injection $\mathbf{Y}$ to the implicit equilibrium block. Similar to $f_{\mathrm{eim}}(\cdot,\cdot)$, we design the decoder-side implicit equilibrium block $f_{\mathrm{dim}}(\cdot,\cdot)$ with two parallel branches and identity shortcuts. The powerful computation capability at the BS can support more complicated operations than the users and thus improve the overall reconstruction performance. Instead of using one standard convolution for input injection, we adopt four consecutive convolutions at the decoder. In addition, the number of feature maps is expanded from 2 to 80. The other settings are identical with the encoder. The sigmoid mapping is adopted as the post-processing function $f_{\mathrm{post}}(\cdot)$.

\subsection{Flexibility} \label{flexibility}
The number of FLOPs is used to measure the time complexity of a learning model. To show the flexible execution of our proposed method, we first evaluate the number of FLOPs of different deep learning components. The number of FLOPs of a fully-connected layer with bias is given by
\begin{equation} \label{FLOP_FC}
F_\mathrm{fc} = 2 I_\mathrm{in} I_\mathrm{out},
\end{equation}
where $I_\mathrm{in}$ and $I_\mathrm{out}$ denote the input dimension and output dimension, respectively.
The number of FLOPs of a convolutional layer with bias is expressed as
\begin{equation} \label{FLOP_Conv}
% F_\mathrm{conv} = (2 \times C_\mathrm{in} \times K^2)\times H \times W \times C_\mathrm{out}
F_\mathrm{conv} = 2 C_\mathrm{in} K^2 H  W  C_\mathrm{out},
\end{equation}
where $K$ is the kernel size. $H$ and $W$ are the height and the width of the output feature map, respectively. $C_\mathrm{in}$ and $C_\mathrm{out}$ denote the numbers of input and output channels, respectively. According to \eqref{FLOP_FC} and \eqref{FLOP_Conv}, the number of FLOPs of each proposed block can be computed. The total number of FLOPs of the proposed encoder $F_\mathrm{e}$ and decoder $F_\mathrm{d}$ is then given by
\begin{equation}
\begin{aligned}
& F_\mathrm{e} = F_\mathrm{pre} + T_\mathrm{e} \times F_\mathrm{eim} + F_\mathrm{down}, \\
& F_\mathrm{d} = F_\mathrm{up} + T_\mathrm{d} \times F_\mathrm{dim} + F_\mathrm{post},
\end{aligned}
\end{equation}
where $F_{i}$ denotes the number of FLOPs of function $f_{i}$, for $i \in \{\mathrm{pre}, \mathrm{eim}, \mathrm{down}, \mathrm{up}, \mathrm{dim}, \mathrm{post}  \}$.

Due to the dynamic communication environment, the available computation capability of devices varies from time to time. Given the computational budget $R_\mathrm{e}$ and $R_\mathrm{d}$ at the users and the BS, respectively, the number of FLOPs of the implemented encoder and decoder should not exceed the budget, i.e., $F_\mathrm{e} \leq R_\mathrm{e}$ and $F_\mathrm{d} \leq R_\mathrm{d}$. However, the $F_\mathrm{e}$ and $F_\mathrm{d}$ of conventional deep learning-based approaches are fixed once trained. Therefore, they lack a performance-efficiency trade-off at runtime and models with different complexity need to be trained, benchmarked, and deployed individually. Once the environment changes, i.e., $R_\mathrm{e}$ and/or $R_\mathrm{d}$ changes, one needs to switch to a larger or smaller model by downloading pre-trained weights, which consumes large memory and time cost. In contrast, in our proposed method, the number of iterations of encoding and decoding equilibrium blocks can be adjusted according to the resource budget, i.e.,
\begin{equation}
\begin{aligned}
& T_\mathrm{e} = \lfloor \frac{R_\mathrm{e} - F_{\mathrm{pre}} - F_{\mathrm{down}}}{F_{\mathrm{eim}}} \rfloor, \\
& T_\mathrm{d} = \lfloor \frac{R_\mathrm{d} - F_{\mathrm{up}} - F_{\mathrm{post}}}{F_{\mathrm{dim}}} \rfloor.
\end{aligned}
\end{equation}
This enables a flexible trade-off between accuracy and complexity. When the resource budget varies, the proposed approach can adaptively select an appropriate iteration number and achieving an online accuracy-efficiency trade-off without re-training or downloading data.

\subsection{Training Strategies}
Since the encoding and decoding blocks share similar structures, we take the encoder side as an example for the ease of illustration. According to \cite{bai2019DEQ, bai2020MDEQ}, we can directly backpropagate the implicit equilibrium block using the Jacobian of $f_{\mathrm{eim}}$ at $\mathbf{M}^{*}$, i.e.,
\begin{equation}\label{SGD}
\begin{aligned}
& \frac{\partial \ell}{\partial \theta} = \frac{\partial \ell}{\partial \mathbf{M}^{*}}(\mathbf{I}-\mathbf{J}_{f_{\mathrm{eim}}}|_{\mathbf{M}^{*}})^{-1}\frac{\partial f_{\mathrm{eim}}(\mathbf{M}^{*}, \mathbf{X})}{\partial \theta}, \\
& \frac{\partial \ell}{\partial \mathbf{X}} = \frac{\partial \ell}{\partial \mathbf{M}^{*}}(\mathbf{I}-\mathbf{J}_{f_{\mathrm{eim}}}|_{\mathbf{M}^{*}})^{-1}\frac{\partial f_{\mathrm{eim}}(\mathbf{M}^{*}, \mathbf{X})}{\partial \mathbf{X}}, 
\end{aligned}
\end{equation}
where $\ell$ denotes the training loss, $\theta$ denotes the trainable parameters, $\mathbf{I}$ is the identity matrix, and $\mathbf{J}_{f_{\mathrm{eim}}}|_{\mathbf{M}^{*}}$ represents the Jacobian of $f_{\mathrm{eim}}$ at $\mathbf{M}^{*}$. Since the computation of the inverse Jacobian is complicated, we further adopt the approximated gradient descent direction in \cite{JFB}
\begin{equation}\label{JFB}
\begin{aligned}
& \hat{\frac{\partial \ell}{\partial \theta}} = \frac{\partial \ell}{\partial \mathbf{M}^{*}}\frac{\partial f_{\mathrm{eim}}(\mathbf{M}^{*}, \mathbf{X})}{\partial \theta}, \\
& \hat{\frac{\partial \ell}{\partial \mathbf{X}}} = \frac{\partial \ell}{\partial \mathbf{M}^{*}}\frac{\partial f_{\mathrm{eim}}(\mathbf{M}^{*}, \mathbf{X})}{\partial \mathbf{X}}.
\end{aligned}
\end{equation}
It is proved in \cite{JFB} that \eqref{JFB} is a descent direction of $\ell$, even for approximate fixed-points. By adopting \eqref{JFB}, we separate the forward fixed-point finding procedures from the backward neural network training. The backpropagation is thus based on differentiating through one layer at the fixed-point, i.e., $\frac{\partial f_{\mathrm{eim}}(\mathbf{M}^{*}, \mathbf{X})}{\partial \theta}$ and $\frac{\partial f_{\mathrm{eim}}(\mathbf{M}^{*}, \mathbf{X})}{\partial \mathbf{X}}$. No intermediate values in the equilibrium block are required and the training memory consumption is constant, which is equivalent to training a one-layer neural network. % The whole algorithm is then trained end-to-end. 

\section{Simulation Results} \label{simulation}
In this section, we demonstrate the performance of the proposed lightweight and flexible CSI feedback approach for FDD massive MIMO systems. % , and compare it to two baselines introduced in Section \ref{existing}.

%\begin{table*}[t]	
%% \selectfont  
%\centering
%\newcommand{\tabincell}[2]{\begin{tabular}{@{}#1@{}}#2\end{tabular}}
%\caption{Performance versus Space Complexity.} 
%\resizebox{1\textwidth}{!}{
%\begin{tabular}{|c|c|c|c|c|}
%\hline
%Compression Ratio & Method & Indoor NMSE & Outdoor NMSE & Parameter Number \\ \hline
%%  \cite{} & \tabincell{c}{
%%Achievable spectral efficiency in \\ single-user MISO systems
%%} & ${\bf{w}}$: Transmit beamforming & ${\left\| {\bf{w}} \right\|^2} \le P $  \\ \hline
%%Secrecy rate maximization \cite{Yu2020Robust} & Sum secrecy rate of $K$ legitimate users & 
%%\tabincell{c}{
%%${\bf{w}}_k$: Transmit beamforming, \\ ${\bf{Z}}$: Artificial noise
%%}
%%& \tabincell{c}{
%%$\sum_{k=1}^{K}\left\|\mathbf{w}_{k}\right\|^{2} + {\rm{Tr}({\bf{Z}})} \leq P$,  \\ ${\bf{Z}} \succeq {\bf{0}}$
%%}   \\ \hline
%%Weighted sum-rate maximization \cite{guo2020weighted} & Weighted sum-rate of $K$ mobile users & ${\bf{w}}_k$: Transmit beamforming & $\sum_{k=1}^{K}\left\|\mathbf{w}_{k}\right\|^{2} \leq P $ \\
%\hline
%\end{tabular}
%}
%\label{example}
%\end{table*}
%% \vspace{-0.5em}

\subsection{Simulation Setup}
\subsubsection{Data Generation}
%Following the experimental setting in \cite{CsiNet}, two types of channel matrices are generated according to the COST 2100 models \cite{COST2100}, i.e., the indoor picocellular scenario working at the 5.3 GHz band and the outdoor rural scenario working at the 300 MHz band. The BS is equipped with the uniform linear array with $N_t = 32$ and the number of subcarriers is 1024. The original $2 \times 1024 \times 32$ CSI matrix is transformed into the angular-delay domain and truncated to the first 32 rows, forming the $2 \times 32 \times 32$ matrix $\mathbf{H}''$.
Following the experimental settings in \cite{CsiNet}, an indoor picocellular system operating at the 5.3 GHz band is considered. The channel matrices are generated according to the COST 2100 models \cite{COST2100}. The BS is equipped with a uniform linear array with $N_t = 32$. The number of subcarriers is set as 1024. The original $2 \times 1024 \times 32$ CSI matrix is transformed into the angular-delay domain and truncated to the first 32 rows, forming the $2 \times 32 \times 32$ matrix $\mathbf{H}''$.

\subsubsection{Training Settings}
The training, validation, and test datasets contain 100,000, 30,000, and 20,000 samples, respectively. The Adam optimizer is used for trainable weight updates. Kaiming initialization is used for each convolution operation and fully-connected layer. The numbers of epochs is 1000 and batch size is set to 200. During training, $T_{\mathrm{e}}$ and $T_{\mathrm{d}}$ are set to 10 and 5, respectively. The training loss is the MSE between the recovered CSI and the true CSI. %The initial learning rate is 0.001 and 
The learning rate varies between $\eta_{\mathrm{min}} = 5\times10^{-5}$ and $\eta_{\mathrm{max}} = 0.001$, and the process can be expressed as
\begin{equation}
\eta_t=\eta_{\min }+\frac{1}{2}\left(\eta_{\max }-\eta_{\min }\right)\left(1+\cos \left(\frac{t}{T} \pi\right)\right),
\end{equation}
where $\eta_t$ is the learning rate of the $t$-th epoch and $T$ denotes total number of epochs.
\subsubsection{Evaluation Metric}
The normalized mean squared error (NMSE) between the recovered channel and the true channel is used to evaluate the performance, which is given by
\begin{equation}
\text{NMSE} = \mathbb{E} \left\{\frac{||\mathbf{H}'' - \hat{\mathbf{H}}''||_2^2}{||\mathbf{H}''||_2^2}  \right\}.
\end{equation}
In addition, the number of FLOPs is used to measure the time complexity of the learning model, and the number of trainable parameters is adopted as a metric to measure the space complexity \cite{ConvCsiNet}. All the simulations are done using the existing deep learning platform PyTorch. The number of FLOPs and trainable parameters are calculated using the thop package \cite{Thop} for PyTorch.

\subsection{Performance Comparison}
To illustrate the effectiveness of the proposed CSI feedback design, we adopt three benchmarks for comparison:
\begin{itemize}
%\item \textbf{WMMSE Perfect CSIT}: Assuming perfect downlink CSI at the transmitter (CSIT), the conventional iterative WMMSE method in \cite{Shi11WMMSE} is used for beamforming. This baseline serves as a performance upper bound with a high computational complexity.
%\item \textbf{ZF Perfect CSIT}: The ZF beamforming solution is employed given perfect CSIT.
\item \textbf{CsiNet \cite{CsiNet}}: An exploratory work that enjoys low time and space complexity.
\item \textbf{ConvCsiNet \cite{ConvCsiNet}}: A complicated CNN-based method that achieves good performance but induces heavy computational costs.
\item \textbf{ShuffleCsiNet \cite{ConvCsiNet}}: An efficient neural network architecture is adopted, but the complexity is still high for users with extremely limited computational power.
\end{itemize}

%%%%%%%%%%%%%%%%%%%%%%% Fig1
\begin{figure}[t] %H为当前位置，!htb为忽略美学标准，htbp为浮动图形
\centering
\includegraphics[height=5.4cm]{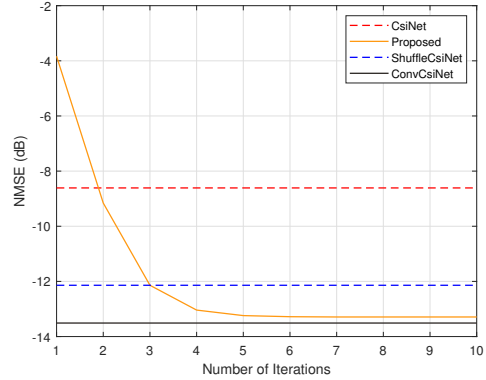} %插入图片，[]中设置图片大小，{}中是图片文件名
\caption{NMSE achieved by different methods versus $T_\mathrm{e}$ when $\gamma = 1/16$ and $T_\mathrm{d}=5$.} %最终文档中希望显示的图片标题
\label{convergence} %用于文内引用的标签
\end{figure}

Fig. \ref{convergence} plots the NMSE achieved by the proposed deep equilibrium learning-based scheme and other baselines versus the number of iterations when $\gamma = 1/16$. Since the other baselines use explicit neural networks and the outputs are acquired through one forward propagation, they do not have the concept of iterations. It can be observed from Fig. \ref{convergence} that running the proposed method with just 2 iterations outperforms CsiNet and 3 iterations outperforms ShuffleCsiNet, showing the superiority of the proposed scheme in terms of efficiency. Moreover, it is demonstrated that the proposed CSI feedback scheme converges with 7 iterations and achieves a comparable performance as ConvCsiNet.  

\begin{figure}[t] %H为当前位置，!htb为忽略美学标准，htbp为浮动图形
\centering
\includegraphics[height=5.4cm]{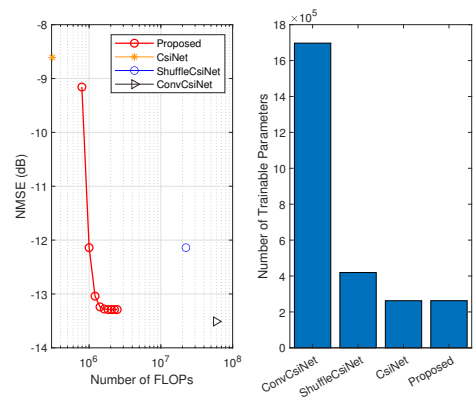} %插入图片，[]中设置图片大小，{}中是图片文件名
\caption{Performance and complexity of different methods when $\gamma = 1/16$. (Left) NMSE achieved by different methods versus the number of FLOPs at the encoder side. (Right) The number of trainable parameters of different methods at the encoder side.} %最终文档中希望显示的图片标题
\label{tradeoff} %用于文内引用的标签
\end{figure}

%\begin{figure} \centering    
%\subfigure[$\gamma = 1/16$] {
% \label{fig:a}     
%\includegraphics[width=0.77\columnwidth]{CR1_16_subfigure_crop.pdf}  
%}     
%\subfigure[$\gamma = 1/32$] { 
%\label{fig:b}     
%\includegraphics[width=0.77\columnwidth]{CR1_32_subfigure_crop.pdf}     
%}    
%\caption{Accuracy-efficiency trade-offs of different methods. (Left) NMSE achieved by different methods versus the number of FLOPs at the encoder side. (Right) The number of trainable parameters of different methods at the encoder side.}     
%\label{tradeoff}     
%\end{figure}

In Fig. \ref{tradeoff}, we demonstrate the accuracy-efficiency trade-offs of different methods. In the left sub-figure, the NMSE achieved by different methods versus the number of FLOPs at the encoder side is demonstrated. It is shown that although CsiNet only requires a small number of FLOPs and induces low time complexity, the performance is not satisfactory. On the other hand, the ConvCsiNet method achieves around 5 dB higher accuracy compared with CsiNet, but it requires more than 58 million FLOPs. Assume that the compression ratio is $1/16$ and the CSI feedback and recovery period is 1 millisecond. The computational power required by the ConvCsiNet encoder is about 59 G floating point operations per second (FLOPS) \cite{ConvCsiNet}. Note that Kirin 970, one of the mid- and high-end mobile systems on chip (SoC), has a total peak computation capability of 244.8 G FLOPS \cite{SoC}. If the ConvCsiNet is deployed in practice, around $1/4$ of the mobile's computational power is used for CSI feedback, making other computationally-intensive tasks such as graphics rendering and voice recognition unable to work. 

In the left sub-figure of Fig. \ref{tradeoff}, it is also demonstrated that the proposed deep equilibrium learning-based CSI feedback design can achieve an instant accuracy-efficiency trade-off by adjusting the encoder-side equilibrium block iteration number at runtime, while for other baselines, the learning model is fixed once trained. Besides, after convergence, the performance of the deep equilibrium learning-based approach is comparable with ConvCsiNet while the number of FLOPs is greatly reduced.

In the right sub-figure of Fig. \ref{tradeoff}, the numbers of trainable parameters of different methods at the user side are plotted. Thanks to the weight sharing in the equilibrium block, the number of trainable parameters is greatly reduced, leading to a lower space complexity of the proposed scheme. It is also demonstrated that the ConvCsiNet has the highest space complexity among all the benchmarks. For example, when $\gamma = 1/16$, the ConvCsiNet encoder network needs to store more than 1 million floating point data, occupying about 6 MB of storage space \cite{ConvCsiNet}. In contrast, the proposed scheme only requires less than 0.27 million float data and thus less than 1.62 MB storage space. This verifies the high efficiency of the proposed CSI feedback design.

%\begin{figure}[t] %H为当前位置，!htb为忽略美学标准，htbp为浮动图形
%\centering
%\includegraphics[height=5.4cm]{NMSEvsCR.eps} %插入图片，[]中设置图片大小，{}中是图片文件名
%% \includegraphics[width=0.4\textwidth]{PerformanceVSantennas_8user.eps} 
%% \includegraphics[width=0.4\textwidth]{NMSEvsLayers_new_crop.pdf}
%\caption{NMSE achieved by different methods versus compression ratios when $T_\mathrm{e}=10$ and $T_\mathrm{d}=5$.} %最终文档中希望显示的图片标题
%\label{NMSEvsCR} %用于文内引用的标签
%\end{figure}

\begin{table}[t]
\begin{centering}
\caption{Performance and Encoder FLOPs for Different Compression Ratios when $T_\mathrm{e} = 10$ and $T_\mathrm{d}=5$ \label{NMSEvsCR}}
\par\end{centering}
\centering{}%
% \begin{tabular}{c|c|>{\centering}p{0.8cm}|>{\centering}p{0.7cm}}
\begin{tabular}{|c|c|c|c|}
\hline 
Compression Ratio $\gamma$& Method & NMSE (dB) & FLOPs (M)\tabularnewline
\hline 
\multirow{4}{*}{$1/4$} & ConvCsiNet & -15.13 & 60.69\tabularnewline
\cline{2-4} \cline{3-4} \cline{4-4}
 & ShuffleCsiNet & -17.36 & 24.11\tabularnewline
\cline{2-4} \cline{3-4} \cline{4-4}
 & CsiNet & -17.36 & 1.09\tabularnewline
\cline{2-4} \cline{3-4} \cline{4-4}
 & \textbf{Proposed} & -19.82 & 3.25\tabularnewline
\hline 
\multirow{4}{*}{$1/8$} & ConvCsiNet & -14.38 & 59.51\tabularnewline
\cline{2-4} \cline{3-4} \cline{4-4}
 & ShuffleCsiNet & -14.59 & 22.93\tabularnewline
\cline{2-4} \cline{3-4} \cline{4-4}
 & CsiNet & -13.47 & 0.57\tabularnewline
\cline{2-4} \cline{3-4} \cline{4-4}
 & \textbf{Proposed} & -15.30 & 2.73\tabularnewline
\hline 
\multirow{4}{*}{$1/16$} & ConvCsiNet & -13.51 & 58.92\tabularnewline
\cline{2-4} \cline{3-4} \cline{4-4}
 & ShuffleCsiNet & -12.14 & 22.34\tabularnewline
\cline{2-4} \cline{3-4} \cline{4-4}
 & CsiNet & -8.65 & 0.31\tabularnewline
\cline{2-4} \cline{3-4} \cline{4-4}
 & \textbf{Proposed} & -13.29 & 2.46\tabularnewline
\hline 
\multirow{4}{*}{$1/32$} & ConvCsiNet & -10.34 & 58.62\tabularnewline
\cline{2-4} \cline{3-4} \cline{4-4}
 & ShuffleCsiNet & -9.41 & 22.05\tabularnewline
\cline{2-4} \cline{3-4} \cline{4-4}
 & CsiNet & -6.24 & 0.18\tabularnewline
\cline{2-4} \cline{3-4} \cline{4-4}
 & \textbf{Proposed} & -9.31 & 2.33\tabularnewline
\hline 
\end{tabular}
\end{table}
Table \ref{NMSEvsCR} presents the FLOPs and NMSE versus the compression ratios $\gamma$. It is demonstrated that ConvCsiNet method achieves the best performance when $\gamma$ is small but the performance degrades when the compression ratio is large. On the other hand, CsiNet and ShuffleCsiNet work well when $\gamma$ is large but the NMSE is not satisfactory when $\gamma$ is small. It is also observed that the proposed method obtains a comparable performance as the baselines for all considered $\gamma$ and outperforms all the baselines when $\gamma = 1/4$ and $\gamma = 1/8$ with a significantly reduced number of FLOPs.

\section{Conclusions}
In this paper, we developed a deep equilibrium learning-based model for CSI feedback in FDD massive MIMO systems. In contrast to existing explicit deep neural networks whose output is characterized by successive non-linear layers, we utilized a fixed-point equation to specify the input-output relationship. The proposed approach is lightweight and permits a flexible accuracy-efficiency trade-off at runtime. 
% In contrast to existing explicit deep neural networks where multiple layers are cascaded and the neural network output is characterized by successive operations, we utilized a fixed-point equation to specify the input-output relationship.
%By doing this we provided a novel way to amalgamate domain knowledge with deep learning and the proposed ADU-based method achieves higher spectral efficiency when users are densely distributed compared with fully data-driven method. 
Extensive simulation results demonstrated that the proposed scheme significantly reduces the memory and computational costs without compromising the performance, when compared with existing methods.

\bibliographystyle{IEEEtran}
\bibliography{IEEEabrv,CSIQBF_V6}

\end{document}